\documentstyle[cite,aps,epsfig,amsmath]{revtex}
\twocolumn
\newcommand{\fp}{{f'_{\mathrm{SG}}}}
\newcommand{\Mm}{{M_{\mathrm{min}}}}
\newcommand{\Ts}{{T_{\mathrm{s}}^0}}
\begin{document}

\subsection*{Comment on ``Phase Diagram of the Random Energy Model with
Higher-Order Ferromagnetic Term and Error Correcting Codes due to Sourlas''}

In a recent Letter, Dorlas and Wedagedera (DW) \cite{dw:remfm} have
studied the random energy model (REM) with an additional $p$-spin
ferromagnetic interaction, as a guide to the properties of a $p$-spin
Ising model with both random spin glass and uniform ferromagnetic
exchange, itself relevant to an error-correcting code
\cite{s:ecc}. They showed that the non-glassy ferromagnetic phase,
found for \mbox{$p=2$} \cite{d:rem} to lie between the paramagnetic
and glassy ferromagnetic phases, is squeezed out to larger
ferromagnetic exchange as $p$ is increased and is eliminated in the
limit of \mbox{$p\to\infty$.} Here we note that (i) we have solved the
corresponding problem of a spherical spin system with $p$-spin glass
interactions and $r$-spin ferromagnetic interactions \cite{gs:prsg}
and have shown that for all \mbox{$r\ge{p>2}$} the opposite situation
applies, namely glassy ferromagnetism is suppressed and only
non-glassy ferromagnetism remains, and (ii) a simple mapping yields
the results of DW and generalizations.

The Hamiltonians for both the Ising and spherical
models consist of a disordered and a ferromagnetic term:
\begin{multline}
{\cal H} = \sum_{i_1 < i_2 \ldots < i_p} J_{i_1 \ldots i_p}
\phi_{i_1} \ldots \phi_{i_p} \\ - \frac{J_0 (r-1)!}{N^{r-1}}
\sum_{i_1 < i_2 \ldots < i_r} \phi_{i_1} \ldots \phi_{i_r}\,,
\end{multline}
where the \mbox{$J_{i_1 \ldots i_p}$} are independent Gaussian random
couplings of zero mean and variance \mbox{$p!J^2/2N^{p-1}$,} and
\mbox{$\phi_i^2=1$} for Ising or \mbox{$\frac{1}{N} \sum_i \phi_i^2 =
1$} for spherical spins. The properties of the system can be found
from the free energy \mbox{$f_{\mathrm{SG}}(M)$} of the system with
\mbox{$J_0=0$} and a constrained magnetization $M$. They are obtained
by minimizing the free energy
\begin{align}
f(M) &= f_{\mathrm{SG}}(M) - \frac{1}{r} J_0 M^r,
\label{eqn}
\intertext{with respect to $M$, which means solving}
\fp(M) &\doteq \frac{df_{\mathrm{SG}}(M)}{dM} = J_0 M^{r-1}.
\label{deqn}
\end{align}

Generally $\fp(M)$ is first order in small $M$, diverges as \mbox{$|M|
\to 1$,} and is monotonically increasing in between. For \mbox{$r=1$,}
corresponding to an applied field \mbox{$h=J_0$,} \mbox{$\fp(M)=h$,}
so the equilibrium magnetization increases monotonically with $h$ and
tends to unity as \mbox{$h \to \infty$.} For \mbox{$r=2$,}
\mbox{$\fp(M) = J_0 M$,} so there is always a solution at
\mbox{$M=0$,} and a ferromagnetic solution appears continuously when
\mbox{$J_0 \geq \fp(0)$.} For \mbox{$r>2$,} the transition is to
a magnetization \mbox{$\Mm>0$,} and $\Mm$ increases with $r$.

The true strength of this method is in predicting the onset of
glassiness: this depends on which parts of $f_{\mathrm{SG}}(M)$
correspond to glassy solutions and so varies with model and with
temperature.

In the upper curve of the Figure we show $\fp(M)$ for the REM above
the glass transition temperature $\Ts$; below $\Ts$ the solution is
glassy everywhere. At the temperature shown, the ferromagnetic
transition is to a non-glassy phase for small enough $r$, while for
larger $r$ $\Mm$ is already in the glassy region. As
\mbox{$r\to\infty$,} \mbox{$J_0 M^{r-1}$} approaches a function which
jumps from zero to $J_0$ at \mbox{$M=1$,} so \mbox{$\Mm \to 1$} and
the transition is directly to the glassy ferromagnet.

\begin{figure}
\epsfig{file=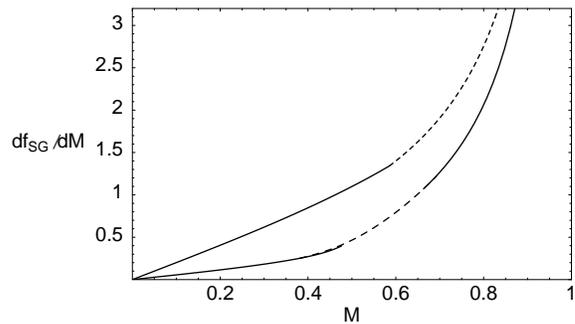,width=3in}
\caption{Plots of \mbox{$df_{\mathrm{SG}}/dM$} versus $M$ for the
random energy model with \mbox{$T/J=0.7$} (upper curve) and for the
spherical $p$-spin model with \mbox{$p=4$} and \mbox{$T/J=0.55$}
(lower curve). The solid lines represent non-glassy solutions, the
dashed glassy.}
\end{figure}

In the lower curve we show $\fp(M)$ for the spherical $p$-spin model
slightly above its $\Ts$; at some higher temperature the glassy region
disappears; below $\Ts$ the glassy region extends down to
\mbox{$M=0$}. At the temperature shown, for small enough $r$, $\Mm$
lies in the lower non-glassy branch, so increasing $J_0$ leads to a
non-glassy ferromagnet, then a glassy, then back to a non-glassy. For
some larger $r$ the first non-glassy ferromagnet disappears, and for
still larger $r$ so does the glassy ferromagnet. A full calculation
\cite{gs:prsg} shows that the second critical value is \mbox{$r=p$}.

The discussion here has been of the static spinodal transition, but it
can be easily extended to the thermodynamic transition by comparing
the free energies (\ref{eqn}) of competing phases; DW concentrate on
this latter case.

\begin{flushleft}
Peter Gillin and David Sherrington\\
Theoretical Physics, 1 Keble Rd, Oxford, OX1\,3NP, UK.\\
\end{flushleft}

\end{document}